\documentstyle[12pt]{article}
\renewcommand{\a}{\alpha}
\renewcommand{\b}{\beta}
\newcommand{\g}{\gamma}
\renewcommand{\d}{\delta}
\newcommand{\D}{\Delta} 

\newcommand{\ep}{\epsilon}
\newcommand{\G}{\Gamma}

\newcommand{\la}{\lambda}
\newcommand{\La}{\Lambda}

\newcommand{\si}{\sigma}
\newcommand{\Si}{\Sigma}

\newcommand{\z}{\zeta}
\newcommand{\zb}{{\bar z}}
\newcommand{\Zb}{{\bar  Z}}

\newcommand{\p}{\partial}
\renewcommand{\pb}{{\bar \p}}
\newcommand{\na}{\nabla}
\def\nab{{\overline  \na}}

\newcommand{\non}{\nonumber\\}

\newcommand{\beq}{\begin{equation}}
\newcommand{\eeq}{\end{equation}}
\newcommand{\beqa}{\begin{eqnarray}}
\newcommand{\eeqa}{\end{eqnarray}}

\newcommand{\refeq}[1]{(\ref{#1})}
\def\half{{\mbox{\small  $\frac{1}{2}$}}}

\newcommand{\prl}[1]{{  Phys. Rev. Lett.} { {#1}}}
\newcommand{\plb}[1]{{  Phys. Lett.} { {#1B}}}
\newcommand{\pla}[1]{{  Phys. Lett.} { {#1A}}}
\newcommand{\npb}[1]{{  Nucl. Phys.} { B{#1}}}

\newcommand{\prep}[1]{{  Phys. Rep.} { {#1}}}

\newcommand{\cmp}[1]{{ Comm. Math. Phys} { {#1}}}

\newcommand{\ijmod}[1]{{  Int. J. Mod. Phys.} { A{#1}}}

\newcommand{\tp}[2]{t_{\;#1}^{+\;\,#2}}
\def\CP{{\cal P}_M}
\def\Xt{{\tilde X}}
\def\CM{{\cal M}}

\def\GM{G^{(M)}}

\def\JP{J^{(P)}}
\def\es{\ep^{(S)}}
\def\esi{\ep^{(\Si)}}

\begin{document}

\baselineskip 20pt plus 2pt
\begin{titlepage}
\renewcommand{\thefootnote}{\fnsymbol{footnote}}
\begin{flushright}
\parbox{1.5in}
{
 UW-IFT-19/96\\
 August, 1996\\
hep-th/9609196\\}
\end{flushright}
\vspace*{.5in}
\begin{centering}
{\Large \bf Rigid string instantons are pseudo-holomorphic curves\footnote{ 
Work supported in part
by Polish State Committee for Scientific Research (KBN) and EC under the
contract ERBCIP-DCT94-0034.
}}\\
\vspace{2cm}
{\large        Jacek Pawe\l czyk}\\
\vspace{.5cm}
        {\sl Institute of Theoretical Physics, Warsaw University,\\
        Ho\.{z}a 69, PL-00-681 Warsaw, Poland.}\\
\vspace{.5in}
\end{centering}
\begin{abstract}
We show how to find explicit expressions for rigid string instantons for
general 4-manifold $M$. It appears that they are
pseudo-holomorphic curves in the twistor space of $M$. We present explicit
formulae for $M=R^4,\;S^4$. 
We  discuss their properties and speculate on relations to topology of
4-manifolds and the 
theory of  Yang-Mills fields.
\end{abstract}
\end{titlepage}

\renewcommand{\thefootnote}{\arabic{footnote}}
\setcounter{section}{-1}
\setcounter{footnote}{0}
\section{Introduction}

Instantons plays an important role in modern field theory and  mathematics. 
Till now the thorough studies of instantons were carried for gauge fields
and sigma models. Rigid string is another model of interest which was known to
posses instantons. The model was originally considered as a
string for gauge fields. Unfortunately, rigidity has quite complicated
structure when expressed in ordinary string variables. This prevents any
significant progress in quantization. Certain rigid string instantons were
derived and investigated in \cite{polrig,wheater}. 
Despite this efforts  little was known about the generality of the proposed
instanton 
equations  and its significance for physics of the model.

In recent paper \cite{inst} we
derived a new set of instanton equations for the 4d rigid string. 
It was claimed  
this  set is rich enough to have representatives for all topological
sectors of the rigid string. Because the action of the model contains terms
with four derivative  the relevant topological invariant is not only the genus of
the world-sheet surface but also the self-intersection number of the surface
immersed in a target 4d space-time. The instantons split not into
two families - instantons and anti-instantons but into three families.  
We shall call them $\JP_1$-instantons, anti-$\JP_1$-instantons and minimal or 
$\JP_2$-instantons.
Minimal instantons are just minimal 
maps from the world-sheet to the target space-time. 
In general, intersection of these 
families is  non-trivial even in $R^4$ what is also a novel feature.
Unfortunately the equations seemed to be very difficult and a
method (through the Gauss map) to solve them in full generality, failed.

In this paper we are going to study the rigid string instantons of \cite{inst}
in more 
general setting. Thus we shall consider the rigid string moving in a
Riemannian 4-manifold $M$ with the metric $\GM_{\mu\nu}$. Using the twistor
method \cite{eells,ward}
we shall be able to show that in many cases one can give explicit formulas for
the instantons. Moreover the construction will reveal an interesting structure
of the equations, namely, the instantons will appear to be pseudo-holomorphic
curves in the twistor space of $M$. This unexpected result unfolds the
underlying simplicity of the equations and lies foundation of the successful
solution of the equations.  

It is worth to note that  the subject touches Yang-Mills
fields in two points. First of all, pseudo-holomorphic curves
were 
used to build the string picture of YM$_2$ \cite{cmr}. Secondly the
dimension of the moduli space of  $\JP_1$-instantons on $R^4$ and
$S^4$ is exactly the same as those of $SU(2)$ Yang-Mills instantons with the
appropriate identification of topological numbers.

Content of the paper is the following: in Sec.\ref{sec:rig} we introduce the
necessary notation and recall some results of \cite{inst}. In the next section
we show how to solve the $\JP_1$-instanton equations using the twistor method.
We also 
calculate the dimension of the moduli space of the rigid string 
instantons. In sec.\ref{sec:examples} we derive explicit formulas for the
cases of $M=R^4$ and $M=S^4$. In the final section we speculate on
new topological (smooth) invariants of 4-manifolds. We also discuss connection
of the rigid string instantons to string description of 
Yang-Mills fields and shortly discuss the case of
3-dimensional target $M$.

\section{Rigid string instantons.}
\label{sec:rig}
In this section we introduce necessary notions and recall basic results of
\cite{inst}. 

We start with some generalities concerning the problem. 
We shall be interested in
maps $X:\Si\to M$ which are immersions i.e. $rank(dX)=2$ (  
the tangent map is of maximal possible rank).  
Roughly speaking it means that the image of $\Si$ in $M$ is smooth.
It means also that the induced metric $g_{ab}\equiv \p_a{\vec X}\p_b{\vec 
X}$ is non-singular.
Any immersion
defines  the Gauss map  $t^{\mu\nu}:\Si\to G_{4,2}=S^2_+\times S^2_-$. 
The appearance of product of two $S^2$ corresponds to the 
fact that $t^{\mu\nu}$ can be decomposed into self-dual $t^{\mu\nu}_+$ and 
anti-self-dual $t^{\mu\nu}_-$ part. 
If $M$ has non-trivial 
topology we can not expect the Gauss map to be defined globally. Thus we must 
introduce the so-call Grassmann fiber bundle over $M$ with fibers $G_{4,2}$.
The map $\Xt$ to this bundle is called the Gauss lift. Because the
fiber of this bundle splits into self-dual and anti-self-dual part we can
consider Gauss lifts to each of them independently i.e. we can define bundle
of tensors $\tp{mu}{nu}$ separately. This is a sphere bundle  which shall play a crucial role in the next section.

The  action of the rigid string (without the Nambu-Goto
term\footnote{The Nambu-Goto term breaks space-time scale invariance of the model
thus prevents existence of instantons.}) is
\beqa
\int_\Si\sqrt{g}g^{ab}\na_a t^{\mu\nu}\na_b
t_{\mu\nu}=
2\int_\Si\sqrt{g}(\D{ X^\mu})(\D{ X^\nu})\GM_{\mu\nu}-8\pi \chi.   
\label{extc1}
\eeqa
where 
$t^{\mu\nu}\equiv \ep^{ab}\p_a X^\mu \p_b X^\nu/\sqrt{g}$ are the element 
of the Grassmann manifold $G_{4,2}$, $\GM_{\mu\nu}$ is the metric on $M$ and
$g_{ab}\equiv  \p_aX^\mu \p_b X^\nu \GM_{\mu\nu}$ is the induced metric on a
Riemann surface of genus $h$.
Tensors $\p_a X^\mu$ are components of 
$T^*\Si\otimes X^*TM$, where  $X^*TM$ is the pull-back
bundle. The covariant derivatives are built with Levi-Civita connections on
$T^*\Si$ and $TM$. Explicitly $\na_b \p_aX^\mu
=\p_b\p_aX^\mu-\G^{(\Si)c}_{\;\;ab} 
\p_cX^\mu+\G^{(M)\mu}_{\;\;\rho\si} \p_aX^\si \p_bX^\rho$.
The Euler
characteristic of the Riemann surface $\Si$ is given by the Gauss-Bonnet formula
$\chi=\frac{1}{4\pi}\int_\Si\sqrt{g}R$.  

Immersions of Riemann surfaces  in $R^4$ are classified by  the
self-intersection number $I$ \cite{whitney}.
General arguments based on singularity theory showed that rigidity separates
topologically different string configurations. 
The derivation of instanton equations  was based on the knowledge of relevant
topological invariants. In our case these were the above mentioned
self-intersection  number $I$ and
the Euler characteristic $\chi$. The equations were derived using
formulae for both invariants in terms of $t^{\mu\nu}$. Explicitly:
$\chi=I_+-I_-, \quad I=\half(I_++I_-)$,
where $I_{\pm}=\pm\frac{1}{32\pi}\int_\Si \ep^{ab}\p_a t^{\mu}_{\pm\;\nu}
\p_b t^{\nu}_{\pm\;\rho} t^{\rho}_{\pm\;\mu}$ and $t_{\pm}^{\mu\nu}\equiv
t^{\mu\nu}\pm {\tilde t}^{\mu\nu}$.
Standard reasoning yielded the following instanton equations
(denoted  as $(+,\pm)$ with obvious sign convention)\footnote{The minus $-$ in
front 
of the first term appeared in order to preserve notation of \cite{inst}.}:
\beqa
-\na_a \tp{\;\mu}{\nu}\pm \frac{\ep_a^{\;\;b}}{\sqrt{g}}
\tp{\;\rho}{\nu} \na_b \tp{\;\mu}{\rho}=0
\label{instpm}
\eeqa
Here the equations were adopted to the general manifold $M$.
Analogous equations hold for the anti-self-dual part of
$t^{\mu\nu}$. The $(+,+)$  equations are equivalent to
$\D X^\mu=0$ and their solutions will be called minimal 
instantons or $\JP_2$-instantons. Appropriate equations for anti-self-dual
part of $t$ will give  
only $\JP_1$-anti-instantons - the fourth possibility appeared to be
equivalent to the minimal instantons.
Thus instantons form 3 families.
The former two families behave as true instantons and anti-instantons in this
sense that they do not have continuation to the Minkowski space-time and their
role is interchanged under change of orientation of the space-time. Minimal
instantons have continuation to Minkowski space-time what is a novel feature
of this kind of solutions. It is also worth to note that change of orientation
of the world-sheet (change of sign of the world-sheet complex structure)
together with change of sign of $t^{\mu\nu}$ 
(change of sign of the space-time complex structure) do not change any of the
equations. 

In $R^4$ the instanton families are not disjoint. 
Intersection of $\JP_1$-instantons and minimal
instantons gives $\na_a \tp{\;\mu}{\nu}=0$ while intersection of
anti-$\JP_1$-instanton and minimal 
instantons gives $ \na_a t^{-\;\,\mu}_{\,\nu}=0$. 
These equations have solutions in $R^4$ \cite{wheater} and $S^4$
\cite{bryant,eells}. 
For $R^4$ there is also one
nontrivial 
intersection of $\JP_1$-instantons and anti-$\JP_1$-instantons at genus zero. The
solution was  
found in \cite{inst} to be a sphere embedded in $R^3\subset R^4$. It  
has 5-dimensional moduli space - four positions and one breathing
mode. The analogy with  SU(2) Yang-Mills case is suggestive. In 
sec.\ref{sec:twistor} we shall show that in fact the dimension of the
moduli space of  $\JP_1$-instantons on $R^4$ and $S^4$ is exactly given by the
same formula as for 
the $SU(2)$ Yang-Mills case with appropriate identification of topological numbers.
We want to stress here that these properties of the three family of instantons
were proven for $M=R^4$ and may be modified for other 4-manifolds.

The above mentioned spherical solution was found using properties of the Gauss
map of an immersion. Unfortunately we were not able to find other instantons
with this method. In the following section we shall use the twistors
\cite{eells,ward} in finding solutions to Eqs.\refeq{instpm}. The method
appears so powerful that 
one can find closed formulae for all instantons for many interesting spaces $M$.

\section{Twistor construction of instantons}
\label{sec:twistor}

In this section we shall show that the rigid string instantons are
pseudo-holomorphic 
curves in the twistor space of the space-time $M$. This will directly lead to
the explicit formulas on instantons for some manifolds $M$. Moreover, the
method will allow to calculate the dimension of the moduli space of
instantons. In the following we shall concentrate on the
self-dual part of $t^{\mu\nu}$ only, understanding that the behavior of the
anti-self-dual part is analogous.

Before we go to the main subject we recall some facts from complex
geometry and twistors.
We shall heavily use certain properties of almost complex
structures.\footnote{Several
different almost complex structure will appear in this paper. In order to
clarify the notation we 
decided to denote by $\ep$ almost complex structures of 2d manifolds 
and by $J$ almost complex structure of $M$ and the twistor space $\CP$. These
will be supplemented with the  
appropriate superscript of the manifold. An almost complex manifold $X$ with
given almost complex structure $J$ we shall denote as $(X,J)$.}

Thus, the space of all almost complex structures
on $R^4$ is $O(4)/U(2)=S^2\times {Z}_2$ i.e. the space of all orthonormal
frames up to unitary rotation which preserve  choice of complex coordinates. 
The $Z_2$ factor is responsible for change of orientation of $R^4$. 
Hence the bundle of almost complex structures over $M$ (up to change of
orientation) is just the  
sphere bundle $\CP$ i.e. a bundle with $S^2$ as fibers. 
In other words, any  point $p\in \CP$, with coordinates in a local
trivialization 
$p=(u,x) \in S^2\times R^4$,  fixes 
an almost complex structure $J$ on $M$ at $x=\pi (p)\in M$. This almost complex
structure is given by the coordinate $u$ on the fiber $S^2$. 

It appears that such sphere bundles have two 
natural almost complex structures. The reason is that 
the sphere $S^2$  has two canonical complex structures $\pm \ep^{(S)}$. In the
conformal metric on $S^2$ we have $(\ep^{(S)})^{ij}=\ep^{ij}$, 
where $i,j=1,2$. Out of $\pm\es$ we build two almost complex structures on $\CP$.
With the help of Levi-Civita connection on $M$ we can decompose the tangent
space $T_p\CP$ at $p\in \CP$ into the horizontal part 
$H_p$ and the vertical part $V_p$: $T_p\CP=H_p\oplus V_p$. 
The former is isomorphic to $T_{\pi (p)}M$. The isomorphism is given
by  the lift defined with help of the Levi-Civita connection on $TM$. The lift
also defines the
action of the almost complex structure $J$ on $H_p$. The vertical space $V_p$
is tangent to the 
fiber ($S^2$) and has the complex structure $\pm \ep^{(S)}$. It follows
that we can define two almost complex structure at $p\in \CP$ given by the
formulas 
\beqa
\JP_1&=&J\oplus \ep^{(S)}\non
\JP_2&=&J\oplus -\ep^{(S)}.
\label{acs}
\eeqa
Both almost complex structures \refeq{acs} will appear in the subsequent
construction of the rigid string instantons. The sphere bundle $\CP$ with given
almost complex structure $\JP_1$ or $\JP_2$ is sometimes called  the twistor
space 
\cite{eells} (see also \cite{ward}).

Now let us recall that a (complex) curve $y$ from a Riemann surface
$(\Si,\ep^{(\Si)})$  to a manifold $(N,J^{(N)})$ is said to be
pseudo-holomorphic if 
\beqa
dy+J^{(N)}\circ dy\circ\esi=0
\label{jholo}
\eeqa
where $dy$ is the tangent map $dy:T\Si\to TN$. Sometimes, in order to
indicate the almost 
complex structure of the target space we shall call \refeq{jholo}
$J$-holomorphic curve suppressing reference to $\esi$ \cite{gromov,dusa}.  
In this paper we shall take 
$\esi$ to be  complex structure given by
$(\esi)_a^{\;\;b}=g_{ac}\ep^{cb}/\sqrt{g}$ where $g$ is the 
metric o n $\Si$. When we pull back the definition \refeq{jholo}
on $\Si$ we get
\beqa
\p_a\, y^m+  (J^{(N)})_n^{\;\;m}\; \p_b\, y^n\; \frac{ \ep_{a}^{\;\;b}}{\sqrt{g}}=0. 
\label{jholoi}
\eeqa
$(a,b,c=1,2\quad m,n=1,...\dim\CP)$. For the conformal metric and complex
coordinates on $\Si$ \refeq{jholoi} is $\pb y^m-i (J^{(N)})_n^{\;\;m}\;\pb y^n  =0$. Thus 
$\half (1-iJ^{(N)})$ is the projector on the holomorphic part, while 
$\half (1+iJ^{(N)})$ on anti-holomorphic part of (complexified) $TN$.

As it was established in the previous section any immersions defines a sphere
bundle. 
Explicitly we define $\CP$ as the bundle of normalized, self-dual  
tensors $t^{\mu\nu}_{+}$
over $M$. The fiber of this bundle is homeomorphic to $S^2$
(the normalization is $t_{+}^{\mu\nu}t_{+\mu\nu}=4$). The Gauss lift to this
bundle will be denoted by $\Xt_+$.
\beqa
\begin{array}{ccc}    &                          &     \CP       \\ 
                     &\stackrel{\Xt_+}{\nearrow} & \downarrow  \pi   \\
                \Si  & \stackrel{X}{\longrightarrow}              & M
                         \end{array} 
\eeqa
We see that 
this bundle is isomorphic to the bundle of almost complex structures defined
previously. This is 
the reason why we used the same notation in both cases.

After establishing this simple fact we go to the instanton equation
\refeq{instpm}. 
We rewrite \refeq{instpm} and the equation which follows from definition of
$t_+$  in the conformal gauge for the induced metric
$g_{ab}\equiv  \p_a X^\mu\p_b X^\nu \GM_{\mu\nu}\propto \d_{ab}$. 
\beqa
(+,\pm)=\nab \tp{\mu}{\nu}\pm i
 \tp{\rho}{\nu}\,\nab \tp{\mu}{\rho}&=&0\non
\pb X^\mu-i\tp{\rho}{\mu}\,\pb X^\rho&=&0
\label{insteq}
\eeqa
We have chosen complex 
coordinates on  $\Si$, thus $\nab$ is the anti-holomorphic part of the covariant
derivative.  
 
Next we show that Eqs.\refeq{insteq} give pseudo-holomorphic curves on $\CP$
with the 
two almost complex structures \refeq{acs}.
Any Gauss lift defines $t_+^{\mu\nu}$ and hence with the
help of the metric  $\GM_{\rho\nu}$ we can write down an expression for the almost 
complex structure $J_{\mu}^{\;\;\nu}= t_{\;\mu}^{+\;\,\nu}$ on $X^* TM$ at $z\in\Si$. 
We emphasize that $J$ depends on coordinates on the Grassmann
bundle $\CP$.
This almost complex structure  decomposes (the complexification of) 
the tangent space $X^*T_{X(z)}M$
into holomorphic $T^{(1,0)}$ and anti-holomorphic $T^{(0,1)}$
part.\footnote{We shall suppress the index $X(z)$ of the tangent space at this
point.}
The former is defined as the 
space of vectors of the form $T^{(1,0)}=\{(1-iJ)V;V\in X^*TM\}$ while the latter  
are complex conjugate vectors.
We also choose locally almost 
hermitian  metric which 
 provides the following identification:
$T^{(0,1)}=T^{*(1,0)}$ and
$T^{*(0,1)}=T^{(1,0)}$. Thus from tautology
$\half(1-iJ)\half(1-iJ)\half(1-iJ)=\half(1-iJ)$ we get $\half(1-iJ)\in
T^{(1,0)} \otimes T^{*(1,0)}$ so $J\in T^{(1,1)}$.
From $J(\nab J)+(\nab J) J=0$ we check that 
$\half(1-iJ)[(1-iJ)\nab J]\half(1+iJ)=(1-iJ)\nab J$  i.e. $(1-iJ)\nab J\in
T^{(1,0)} \wedge T^{*(0,1)}\sim T^{(2,0)}$. Similarly $(1+iJ)\nab J\in
T^{(0,2)}$.  Any self-dual tensors
decomposes into 
direct sum $T^{(2,0)}\oplus T^{(1,1)}\oplus T^{(0,2)}$ in the almost complex
structure defined by $J$. This can easily checked in particular orthonormal 
basis of
  $T^{(1,0)}\oplus T^{(0,1)}$, e.g. $\{e_1,e_2,{\bar e}_1,{\bar e}_2\}$.
In this basis $J=e_1\wedge {\bar e}_1 + e_2\wedge {\bar e}_2$ and the two 
other self-dual tensors  are $e_1\wedge e_2, {\bar e}_1\wedge {\bar e}_2$.
We note that  $\nab J$ is also self-dual. Thus 
 $\nab J \in T^{(2,0)}\oplus T^{(0,2)}$. As an immediate implication we infer 
that $\nab J$ span the tangent space to the space of almost complex structures
at the point $J$. 

Using the above we can built two almost complex structure on the fibers $S^2$.
We define $\es$ to be such an almost complex structure that $T^{(2,0)}$ are 
holomorphic
vectors while $T^{(0,2)}$ are anti-holomorphic 
vectors. The choice $-\es$ would reverse holomorphicity
properties. Thus, $(1-iJ)\nab J$ is holomorphic,  while $(1+iJ)\nab
J$ is anti-holomorphic in the $\es$ 
complex structure. One can easily find an explicit realization of $\es$ for 
$M=R^4$. For $J_0^{\;\;i}\equiv n^i$, ${\vec n}\in S^2$ and the following
coordinate system on $S^2$ 
\beqa
{\vec n}=(\frac{f{\bar f}-1}{1+|f|^2},\;-i\frac{f-{\bar 
f}}{1+|f|^2},\;\frac{f+{\bar f}}{1+|f|^2})
\label{ffunction}
\eeqa
we get
\beqa
(1-iJ)\pb J=0\quad\Rightarrow  \quad\pb  f=0.
\eeqa
The above $\es$ is just standard complex structure on $S^2$.
With the help of $\pm\es$ we can define two  
almost complex structures \refeq{acs} on the fiber bundle $\CP$ just as we did
in the beginning of this section. 

Now it is easy to see that rigid string instantons \refeq{insteq} are
pseudo-holomorphic 
curves $\Xt_+:\Si\to (\CP,\JP_{1,2})$. Take $\JP$ given by  that
of \refeq{acs} 
and $J, \es$ defined as above. Hence
if we split the map $\Xt_+$
into vertical and horizontal components of $T\CP$ then applying the notation
of \refeq{acs} we rewrite \refeq{jholo} as
\beqa
(1-iJ)dX=0, \quad (1\mp i\es)(d\Xt_+)^v=0
\label{split}
\eeqa
In the first equation we have identified the horizontal component of the
pseudo-holomorphic equation with its counterpart on $M$. In the second equation
$(d\Xt_+)^v$ denotes the vertical part of the map i.e. the space of 
$T^{(2,0)}\oplus T^{(0,2)}$ vectors. Thus, accordingly   
$(1\mp i\ep^{(S)})(d\Xt_+)^v=(1\mp i J)\nab J$. Recalling that $J=t_+$,
this implies that \refeq{split} is
equivalent to \refeq{insteq}. 
We conclude that for conformal induced metric $g_{ab}\sim \d_{ab}$ on $\Si$
\begin{center}
{{\it 
 pseudo-holomorphic curves \refeq{jholo}
are  solutions of the 
instanton equations \refeq{insteq}.}}
\end{center}

The above considerations were applied in \cite{eells} in the context of
minimal and conformal harmonic maps 
$X:\Si\to M$. In our present nomenclature these maps are $\JP_2$-holomorphic
curves in $\CP$. The almost complex structure  $\JP_2$ is non-integrable what
makes  pseudo-holomorphic curves on the manifold ($\CP,\JP_2$) hard to explore.
We shall not dwell upon the case any more referring the reader to the
existing reviews \cite{eells2,osserman}. 

On the other hand, the case of  
$\JP_1$-instantons maybe relatively easy. The reason is that in
some cases  the almost
complex structure $\JP_1$ is integrable  thus defines a complex structure
\cite{ahs} on $\CP$. There is a nice geometrical condition under which this
happens. 
It states that $M$ must be a half-conformally flat manifold \cite{ahs,eells}.
A lot of classical 4-manifolds respect this condition. In this
work we shall concentrate on $M=R^4,\;S^4$. The other examples  are
$T^4,\; S^1\times S^3,\; CP^2,\; K3$. Hence for the half-conformally flat $M$
there exists complex coordinates  $\z_i$ on $\CP$ and then  \refeq{jholo} is
simply 
\beqa
\pb \z_i=0
\label{holo}
\eeqa 
Thus  $\JP_1$-instantons are just holomorphic
maps $\Si\to \CP$. Another important fact is that if $\JP_1$ is integrable
then it depends only on the conformal class of the metric $\GM$ on $M$. 
This property gives  $\JP_1$-instantons on $R^4$ if they are  known on $S^4$
because 
$R^4$ is conformally equivalent to $S^4$. The sphere bundle $\CP$ for the
latter is $CP^3$ with unique complex structure being precisely $\JP_1$. 
Following this facts we shall construct 
all  $\JP_1$-instantons for $\Si=S^2$ explicitly in the next section.

There is a remark necessary at this point. We have chosen
to work in the  conformal metric $g_{ab}=e^\phi\d_{ab}$ 
on $\Si$ thus fixing the almost complex structure on $\Si$
from the very beginning. For higher genus surfaces Riemann surfaces $\Si$ this
is not possible globally unless one allows for some singularities of the
metric i.e. vanishing of the conformal factor. In such a case
solutions of the instanton equations will be so called branched 
immersions \cite{eells}. One may try to avoid this working
with the most general  complex structure $\esi$. This causes problems
with the definition of almost complex structures on $\CP$. It is  
because,  for the rigid string, $\esi$ 
is determined by the induced metric from $X$, but not  from $\Xt$.
The problem can be resolved if both metrics are the same what happens for
intersection of $\JP_1$ and $\JP_2$ families.
It appears that if $M=S^4$ then  
all minimal surfaces  respect this condition \cite{bryant}.
 
\subsection{Moduli space}

We define the moduli space $\CM$ of the problem \refeq{insteq}  
as the space of solutions modulo automorphism group of solutions and 
reparameterizations of $\Si$.  
This moduli space is the same as the moduli space of \refeq{jholo}.
One of interesting quantities is the dimension of $\CM$.
Unfortunately, fixing the metric on $\Si$ to
be conformal we have lost control (except the case when $\Si=S^2$) 
over the space of reparameterizations.
Thus we first calculate the dimension of  the space ${\tilde \CM}$ of
solutions of 
\refeq{jholo} with fixed metric and then we shall argue how to correct
formula in order to get $\dim(\CM)$. 

The (virtual) dimension of the moduli space
$\dim {\tilde \CM}$ is expressed through an index of an operator \cite{index} 
The latter is a deformation of \refeq{jholo}:   
$\Xt_++\xi:\Si\to \CP$. 
After short calculations we get the deformation of \refeq{jholo}:
\beqa
[(1-i \JP)\nab\xi
-i \nab_\xi\JP \equiv (1-i \JP)\nab\xi+O(\xi)=0. 
\label{oper}
\eeqa
where $O(\xi)$   denotes terms linear in $\xi$ and  not
containing derivatives of $\xi$.  
The operator in \refeq{oper} acting on $\xi$ is the elliptic (twisted)
operator mapping 
$\Xt_+^*T\CP\to\La^{(0,1)}\Si\otimes \Xt_+^*T\CP$. 
Homotopic deformations of the
$O(\xi)$ part does not change its index
\cite{index,dusa}. Thus we can set it to zero and obtain the Dolbeault operator
$\pb_{J}=(1-i \JP)\pb$. The index is given by general Atiyah-Singer theorem or by
Hirzerbruch-Riemann-Roch theorem.
\beqa
{\rm Index}(\pb_{J})&=&c_1(\Xt_+^*T\CP)+\half \dim_C(\CP) c_1(T\Si)\non
&=&c_1(\Xt_+^*T\CP)+3(1-h).
\label{index}
\eeqa
Thus $\dim_R({\tilde \CM})=2c_1(\Xt_+^*T\CP)+6(1-h)$. 
For $g=0$ the moduli space $\CM$  is ${\tilde \CM}$ divided by the action
of the group of automorphisms of $S^2$ i.e. the M{\"o}bius group. Hence we
obtain $\dim_R(\CM)=2c_1(\Xt_+^*T\CP)$. For higher genus surfaces 
$h>0$ if one assumes that the metric on $\Si_h$ is elementary or induced from
$\CP$ one would get 
\beqa
\dim_R(\CM)=
\dim_R({\tilde \CM})-6(1-h)=2 c_1(\Xt_+^*T\CP). 
\label{dim}
\eeqa 
The result agrees with \cite{gromov} where $\CM$ denotes the space of
unparameterized pseudo-holomorphic curves $\Si\to \CP$.
It is interesting to notice that  the formal expression on $\dim_R(\CM)$ is
independent on the almost complex structure on 
$\CP$. Thus one can use the same formula for both families of instantons
\cite{gromov,cmr}. 

It is known that for $M=S^4$ the sphere bundle is $CP^3$. In this case we can
easily find the dimension of $\CM$ for maps from $\Si=S^2$.   
If the map $S^2\to CP^3$ is given by the degree $k$ polynomials in the
variable $z$ we get $\dim_R(\CM)=2c_1(\Xt_+^*CP^3)=2k c_1(CP^3)=8k$. 

\section{Explicit formulae}
\label{sec:examples}
\subsection{ $M=S^4$}
From now on we shall discuss explicit solutions of the $\JP_1$-instanton equations.
 There is vast literature for the minimal
instanton case \cite{eells} and we are not going to review it here. 

It is known that for $S^4$ the
appropriate twistor space is $CP^3$ which has only one complex structure.
Complex projective space $CP^3$ is defined as
projective subspace 
of $C^4$ i.e. 
$CP^3=C^4/\sim$ where $\sim$ means that we identify $(Z_1,Z_2,Z_3,Z_4)$
and $\la (Z_1,Z_2,Z_3,Z_4)$ for all $0\neq\la\in C$. We can cover $CP^3$ with
four charts $k=1,...4$ for which $Z_k\neq 0$ respectively. In the $k$-th chart
we introduce (inhomogeneous) coordinates: $\z_i\equiv Z_i/Z_k$ ($i\neq k)$.
Eq. \refeq{holo} implies that $\z_i$  are meromorphic functions of $z$ on
$\Si$.  This yields instantons on $\CP$ which next must be projected on $S^4$.
We do this with help of a very convenient representation of $S^4$ as the
quaternionic projective  space \cite{atiyah}.
We recall that quaternions are defined as $q=q^m\si^m$ (m=0,..3),
$\si^m=(1,i,j,k)\equiv (1,i{\vec \si})$\footnote{According to the
standard notation, $i$ on the l.h.s. of this definition denotes the matrix, 
 while on the r.h.s., the imaginary unit. This remark is applicable
whenever we use  quaternions.}  
The space of quaternions is denoted by $H$ and is 
isomorphic to $C^2$. The isomorphism is such that $(Z_1,Z_2,Z_3,Z_4)
\leftrightarrow (Z_1+j Z_2,Z_3+j Z_4)\in H^2$. 
Multiplication and conjugation of quaternions follows from
the above matrix representation. Now we have
\beqa
S^4=HP^1\equiv H^2/\sim
\eeqa
In the above  $\sim$ means that we
identify $(q_1,q_2)$ 
and $(q_1q,q_2q)$ for all $0\neq q\in H$ i.e. $S^4$ is quaternionic projective
space (line). 
Quaternionic representation of $S^4$ is so useful because $CP^3$ is complex
projective space in the same $C^4$. 
Heaving a curve in $CP^3$ we can represent it in $H^2=C^4$ and then define two
maps $H^2\to R^4$ which cover $S^4$: $(q_1,q_2)\to (q_1,X_+q_1)$ for
$|q_1|\neq 0$, 
and $(q_1,q_2)\to (X_-q_2,q_2)$ for $|q_2| \neq 0$.
The maps are stereographic projections of $S^4$ from the north and south poles
with the transition function $X_-=1/X_+$. The norm is $|X|^2= 
(X^\dagger X)=X X^\dagger$ (the expression is proportional to the unit
matrix). Explicitly we have
\beqa
X_+=(Z_3+jZ_4)(Z_1+jZ_2)^{-1}=
\frac{(\Zb_1Z_3+Z_2\Zb_4)+j(\Zb_1Z_4-Z_2\Zb_3)}{|Z_1|^2+|Z_2|^2}
\label{quat}
\eeqa  
Rotations $SO(4)=SU_L(2)\times
SU_R(2)/Z_2$ act as $X_+'=(\a_L+j \b_L)X_+(\a_R+j \b_R)$.
We see that the action of both $SU(2)$ groups (here unit quaternions) is
equivalent. 

After these general remarks we go to the detailed description of the 
$\JP_1$-instantons with topology of sphere $S^2$. 
Let us first reproduce the only
compact $\JP_1$-instanton found in \cite{inst}.    
We take $Z_i=a_i (z+b_i)$ (i=1,...4) i.e.
a complex line in $C^4$. For generic choice of $\{a_1,a_2,b_1,b_2\}$
the quaternion $q_1$ is not singular $q_1=Z_1+jZ_2\neq
0$.
By the conformal transformation (M\"{o}bius group), 
$z\to \frac{\a z+\b}{\g z+ \d}$ ($\a,\b,\g,\d\in C,\;\a\d-\b\g=1$), we can
fix position of 3 point. 
Thus we choose $b_1=\infty, b_2=0, a_1=a_2$. Going from $C^4$ to $CP^3$ fixes
$a_1=a_2=1$ so
$(Z_1+j Z_2)=(1+j z)$. Then we get 
\beqa
X_+=\frac{(Z_3+jZ_4)(1-\zb j)}{1+|z|^2}=
X_0+ \frac{(Az+B)+j(-{\bar B}+{\bar A})}{1+|z|^2} 
\label{s4}
\eeqa
for some constants $X_0\in H,\;A,B\in C$. Moding out by the rotation group
leaves only the scale $\la$ and the position $X_0$ as moduli . Hence 
\beqa
X-X_0=\frac{\la}{1+|z|^2}(z+j),\quad \la\in R
\label{sphere}
\eeqa
what is exactly the result obtained in \cite{inst}. \refeq{sphere}
represents sphere of radius $\la/2$. 
The above shows that \refeq{sphere} is the most general
$\JP_1$-instanton 
with $\chi=2,I=0$. 

We can easily generalize this to other topological sectors. 
In order to get $\JP_1$-instantons of the $k$-th sector the
functions $Z_i$ which defines $\z_i$ must be polynomials of degree $k$
\beqa
Z_i=a_i\prod_{j=1}^k (z-a_{ij})\quad i=1,...4
\label{zes}
\eeqa
Thus $\z_i$'s are rational functions with poles at points where coordinates
are ill defined. We can calculate
dimension of the moduli space $\CM$ directly from (\ref{quat},\ref{zes}).
The are $8k+6$ parameters involved in \refeq{quat}. Moding out by the
M{\"o}bius group subtract 6 
parameters yielding $\dim(\CM)=8k$. We can also divide by the rotation group
dropping additional 3 dimensions of the moduli space.
The instanton sectors are characterized by the self-intersection number of the
immersed surface in $S^4$: $I=k-1$.
We shall obtain this result by simple means in the next subsection.
The dimension of the moduli space is
quite remarkable result, because it is exactly the dimension of the moduli space
of $SU(2)$ instantons \cite{atiyah}. 
Moreover we for $k=1$ topology of both spaces is exactly
the same. Topology of $\CM$ for higher $k$ remains to be investigated.

\subsection{$M=R^4$}

It appeared that the rigid string instanton equations, which seemed so
complicated \cite{inst},  can be trivially solved in $R^4$. Using
the parameterization of 
\refeq{ffunction} we can rewrite the second of Eqs.\refeq{insteq} as:
\beqa
\pb {\bar X}_+^1+ f\pb X_+^2&=&0\non
- f \pb X_+^1+\pb {\bar X}_+^2&=&0
\label{flat}
\eeqa
where $X_+^1=X^0+iX^1,\;X_+^2=X^2+iX^3$. This is enormous and unexpected
simplification of the $(1-it_+)\pb X=0$ equation. 
The first line of Eqs.\refeq{insteq} is
\beqa
\pb f&=&0\quad\mbox{ for $\JP_1$-instantons}\\
\pb {\bar f}&=&0\quad\mbox{ for minimal instantons}
\eeqa
Both system of equations are very simple and can be directly integrated.
$\JP_1$-instantons are identical with \refeq{s4}. Explicitly
\beq
X^1_+=\frac{{\bar w}_1(\zb)-{\bar f}(\zb)  w_2(z)}{1+|f|^2},\quad
X^2_+=\frac{{\bar w}_2(\zb)+{\bar f}(\zb)  w_1(z)}{1+|f|^2}
\eeq
Comparing with \refeq{quat} we see that $f= Z_2/ Z_1$. Because
$I_+$ is minus degree of the map:  
$f:\Si\to S^2$ we get $I_+=k$. From the relation: $I_+=I+\chi/2$ follows that
$I=k-1$, what is the result quoted in the previous subsection.

We want to stress that results on $M=R^4$ and $M=S^4$ are almost identical
because $S^4$ is 
conformally equivalent to $R^4$ and the integrable $J_1$ almost complex
structure is conformally 
invariant \cite{ahs}.  We also notice non-triviality of the complex structure
given by $f=f(z)$: holomorphic functions are ${\bar X}_+^1+f\,X_+^2$
and $- f\,X_+^1+{\bar X}_+^2$. 

Minimal instantons can also be integrated and as one could expect they give
solutions of the equation $\D X^\mu=0$. Contrary to the previous case they do
not correspond to minimal surfaces on $S^4$. We shall not dwell upon this
subject referring to the rich existing literature
\cite{eells,bryant,osserman}.  

\section{Speculations and final remarks}

In this section we allude on some possible applications of the presented
results to topology of 4-manifolds and indicate similarities with several
proposals for string picture of gauge fields. We also shortly discuss the case
of 3d target manifold.

\subsection{Topology of 4-manifolds}

Starting from works of Gromov \cite{gromov} and Witten \cite{witten} 
pseudo-holomorphic curves were used to define certain topological invariants, so
called Gromov-Witten invariants \cite{dusa} of symplectic
manifolds (here denoted by $N$). 
The invariants can be defined geometrically in descriptive way as follows:
take a set of homology cycles $\a_i\in H_{d_i}(N,Z)$  and
count (with an appropriate sign) those pseudo-holomorphic curves representing
2-cycle $A\in H_{2}(N,Z)$ which intersect all 
classes ${\a_i}$ at some points. There is also ``physicist'' definition of the
invariants through a correlation function in a topological field theory
\cite{witten}. In
this case the invariants can be formally defined on any almost
complex manifold.

All of twistor spaces 
are almost complex and some of them are K\"ahler (for $M=S^4,\;CP^2$) so also
symplectic. 
Thus following these definitions one could define appropriate invariants for
the twistor spaces of 4-manifolds $M$ considered in this work. 
The hypothesis is:
{\it the Gromov-Witten invariants of the twistor space $\CP$ define some 
invariants of the 4-manifold $M$}.
These new  invariants are well
defined on $M$ if they are well defined on $\CP$. Moreover we can define two
sets of invariants (if we require that $\CP$ must be almost complex only) due
to two natural almost complex structures $\JP_{1,2}$ on $\CP$.

The real problem is what kind of
topological information do they carry? Intersection of cycles in the twistor space
(say at $p\in\CP$) corresponds to the situation when projection of the cycles
to $M$
have common tangents at common point $\pi(p)\in M$. This property is invariant
only under diffeomorphisms of $M$ (class $C^1(M)$) but not under
homeomorphisms of $M$!. It may be that the invariants carry some information
about smooth structures of $M$, so would be similar in nature to Donaldson or
Seiber-Witten invariants. The basic difference is that they are defined in
purely geometrical way avoiding any reference to gauge fields. Moreover the
invariants seems to be well defined on  manifolds for which there are
no other  invariants.
This includes very interesting cases
$M=R^4,\;S^4$ discussed in this paper. Both cases are, of course, different
because there are no compact $\JP_2$-instantons on $R^4$. Contrary, 
the $\JP_1$-invariants should be the same due to one-to-one
correspondence between spaces of instantons in both cases.
This subject, if relevant, seems to be very exciting.

\subsection{Relation to gauge fields}

Going back to physics we want to discuss striking relations of 
rigid string  with gauge fields. Of course both theories uses
twistors in construction of instantons. Leaving this aside we go to more
quantitative 
comparisons. First of all, 
two-dimensional  pseudo-holomorphic curves were
used to build  the string picture of YM$_2$ \cite{cmr}. Rigid string 
instantons provides natural generalization of these curves to 4-dimensions.
One can perform a 
naive 
dimensional reduction of 4d instantons to 2-dimensions by
suppressing two coordinates (say $X^2,\;X^3$). 
This results in taking $|t^{01}|=1$ (there is no distinction between $t_-$
and $t_+$). Thus we get two families of pseudo-holomorphic curves
\beq
\frac{\ep_a^{\;\;b}}{\sqrt{\det (g)}}\,\p_b X^\mu\pm i J_\rho^{\;\;\mu}\, \p_a
X^\rho=0 
\label{twodim}
\eeq
where now $J_\mu^{\;\;\nu}=G_{\mu\rho}\frac{\ep^{\rho\nu}}{\sqrt{\det (G)}}$ and
$G$ is   
the metric on $M^2$. These are  the maps of \cite{cmr} (here $g_{ab}$
is the elementary metric).
On this basis one can state a bold hypothesis that $YM_4$ is localized on the rigid
string instantons\footnote{This is a natural generalization of
\cite{horava}.}. 
All these similarities suggest that rigid string instantons will play a
significant  role in string description of YM fields. Some other ideas along this
line were posed in \cite{nfold}.

We also notice strange coincidence of the
dimensions of the  moduli space of genus zero rigid string instantons on $R^4$ and
$S^4$  and the moduli space of $SU(2)$ Yang-Mills instantons (with the
appropriate identification of topological numbers). For $k=1$ both  moduli
spaces are identical. We do not know what happens for other $k$. 

\subsection{3d manifolds}

Finally we  comment on 3d target manifolds. In this case the tensor
$t^{\mu\nu}$ has 3 components. Classification of immersions of 
surfaces in $R^3$ is more complicated then for the $R^4$ case. There are $4^h$ 
distinct regular homotopy classes of immersions of a surface of genus $h$
into $R^3$ \cite{jamesthomas}. 
One can easily derive appropriate instanton equations following \cite{inst} and
using  $\chi$ only (the self-intersection number $I$ is strictly 4d notion!).
The equations are just Eqs.\refeq{insteq} with $t^{\mu\nu}$ in place of 
$\tp{\mu}{\nu}$. One of the equation is equivalent to $\D X^\mu=0$ another
one represents so-called totally umbilic maps. In the case of $\Si=S^2$ we
have an immediate solution of the latter. This is just the
sphere embedded in $R^3$ given by Eq.\refeq{sphere}. 
Unfortunately, because classification of immersions is so different and we do
not know the invariant which would distinguish all topological classes it is
hard to imagine that the instantons will represent all of them. 
\vskip.5cm

{\bf Acknowledgment}. I would like to thank  Erwin Schr\"odinger Institute 
for kind hospitality where a part of this paper was prepared. 
I also thank E.Corrigan, K.Gaw{\c e}dzki, H.Grosse, B.Jonson, C.Klimcik, A.Morozov,
R.Ward and J.Zakrzewski
for comments and interest in the work. Special thanks to P. Nurowski for many
illuminating discussions on twistor space.

\end{document}